\documentclass[fleqn,twoside]{article}
\usepackage{epsfig}
\usepackage{graphicx}
\usepackage{amssymb}
\usepackage{espcrc2}

\newcommand{\AmS}{{\protect\the\textfont2
  A\kern-.1667em\lower.5ex\hbox{M}\kern-.125emS}}

\title{\large{\bf 
Spectroscopy, leptonic decays and the nature of heavy quarkonia}
\thanks{Research under grant FPA2005-01678; 
Report number IFIC-08-26, FTUV/08-0519.}}
\author{Juan-Luis Domenech-Garret$^a$ and
Miguel-Angel Sanchis-Lozano$^b$ \thanks{Email:Miguel.Angel.Sanchis@uv.es}
\vspace{0.4cm}\\
$^a$Departamento MACS, F\'{\i}sica Aplicada. Universitat de Lleida \\
Alcalde Rovira Roure 191, 25198 Lleida (Spain)\\
$^b$Instituto de F\'{\i}sica
Corpuscular (IFIC) and Departamento de F\'{\i}sica Te\'orica,\\
Centro Mixto Universitat de Val\`encia-CSIC \\
Dr. Moliner 50, E-46100 Burjassot, Valencia (Spain)}

\begin{document} 

\begin{abstract}

We examine the electronic width ratios of 
$\Upsilon$ resonances below the $B\bar{B}$ 
threshold by means of an effective (Cornell-type) QCD potential
incorporating $1/m_b$ corrections obtained from a prior fit
to the bottomonium spectrum. From our analysis we conclude that 
the $\Upsilon(2S)$ and $\Upsilon(3S)$ states should
belong to the strong-coupling (nonperturbative) regime
while the $\Upsilon(1S)$ state should belong to the 
weak-coupling (perturbative) regime, in agreement with a previous study 
based on radiative decays.
\end{abstract}
\maketitle

\section{Introduction}

Heavy quarkonium has historically played a role of utmost importance in the
rise of the Standard Model (SM), notably regarding the quark model
of hadrons and 
the development of QCD as the presently accepted theory to describe the
strong interaction among them. 

Furthermore,  
a large amount of data have been collected 
during the last decade at BEPC, B-factories, CESR, 
HERA, and Tevatron experiments, greatly improving the
accuracy of the measured production cross sections, decay
widths and branching fractions involving heavy quarkonia
(see \cite{Brambilla:2004wf} for a review). In the future, a 
Super Flavour Factory could provide further
experimental results on heavy quarkonia
to an unprecedented accuracy \cite{Bona:2007qt,Akeroyd:2004mj}.
 
On the other hand, 
such precise measurements
are matched by the ever-growing soundness of the theoretical background, firmly
based on an effective field theory, namely the Non-Relativistic QCD
(NRQCD). 
The maturity already reached
in the field even makes feasible the
search for new physics, e.g. in quarkonium decays, looking for experimental
deviations from the SM expectations. 
Let us mention the seek of light dark matter
in invisible quarkonium decays \cite{McElrath:2005bp}
(followed up by experimental searches \cite{Tajima:2006nc,Rubin:2006gc})
and $\Upsilon$ radiative decays into dileptons as
a way of searching for a light non-standard  
Higgs boson \cite{Sanchis-Lozano:2003ha,Dermisek:2006py,SanchisLozano:2006gx,Fullana:2007uq,Mangano:2007gi}.

With the advent of the quark model and QCD, hadronic properties
have been traditionally understood with the help of (more or less
QCD-motivated) potential models, some of them having reached
a fairly acceptable level in predicting
or postdicting level spacings, transitions rates, etc. Nevertheless, 
potential models have several setbacks and limitations, mainly due to the
fact that they do not come directly from first principles.
At this stage, effective theories enter the game
in order to describe rigorously the hadron dynamics.

A prototype is the Heavy Quark Effective Theory (HQET) which 
naturally describes
hadrons with a single heavy quark \cite{Neubert:1993mb}. 
These systems are characterized
by two energy scales: the heavy quark mass, $m_Q$, and 
the characteristic scale of the strong interaction $\Lambda_{QCD}$.
HQET is obtained by integrating out the scale $m_Q$ and
expanding the QCD Lagrangian in powers of $\Lambda_{QCD}$. 

On the other hand, bound states
made of two heavy quarks are characterized by
more scales whose 
relevance and hierarchy are usually estimated by invoking
the so-called velocity counting rules \cite{Lepage:1992tx}. Since the
heavy quark relative velocity $v$ is typically
small ($v^2 \sim 0.3$ for charmonium and $v^2 \sim 0.1$
for bottomonium) the different scales obey the useful relation
$E_n \sim m_Qv^2<<p \sim m_Qv << m_Q$, where $E_n$ is the 
heavy-quark bound-state energy with $n$ the principal quantum number.
NRQCD is obtained by integrating out the heavy quark mass $m_Q$ 
\cite{Bodwin:1994jh}.
High-energy modes
are not lost but encoded into short-distance coefficients and
new local interaction terms in the effective Lagrangian.

The resulting framework allows the separation between 
the short-distance
scale of the process under study from the longer distance scales
associated with quarkonium structure. Therefore operators can
be expanded as a double series (perturbative and nonperturbative), 
with $\alpha_s$ and $v$ being the 
expansion parameters controlling the accuracy of the truncated
series, respectively.
However, these expansion parameters are not completely independent
in heavy quarkonium physics as the typical velocity of
the heavy quark is determined by a balance between the kinetic
energy $m_Qv^2$ and the potential energy which should be
dominated by a Coulombic-term $\sim \alpha_s/r$. Setting 
$r \sim 1/m_Qv$, by invoking the virial theorem we
are led to the well known relation
\[ \alpha_s(m_Qv) \sim v \]

Moreover, since $\alpha_s(\mu)$ runs (decreasing) with the energy scale
(for higher $\mu$), one can approximately write 
\[ \alpha_s(m_b) \sim v^2 \]
where $m_b$ denotes the bottom quark mass. The above relation should play
an important role in assessing the velocity counting rules to be applied
in quarkonium physics. In fact, relativistic corrections of order $v^2$
would be of the same order as perturbative corrections of order 
$\alpha_s(m_b)$.
Let us also remark that the $\alpha_s$ 
perturbative expansions in NRQCD may be not fast convergent
series and truncation even at NNL will likely imply
sizeable effects \cite{Kniehl:1999mx}. In this work,
we will implicitly take the $\alpha_s$ corrections into account 
through the velocity counting rules.

\section{Potential model connection to pNRQCD} 

A naive connection of NRQCD with
potential models can be made by realizing that certain
(long-distance, colour-singlet) matrix elements of NRQCD
can be actually related in first approximation 
to wave functions at the origin (WFO) or their derivatives. 
Nevertheless, this simple picture does not hold too far as 
there are NRQCD matrix elements without an equivalence
in potential models, namely, colour-octet contributions since the
heavy quark antiquark pair needs not to be in a colour-singlet state. 
Actually, the potential picture that emerges from NRQCD 
\cite{Brambilla:2000gk} is quite different from the traditional one 
\cite{Eichten:1995ch,Gonzalez:2003gx} and superior. 

The observation that NRQCD still contains energy scales
irrelevant for the lower-lying states of quarkonium, led
to further simplifications and the resulting theory
was called potential NRQCD (pNRQCD) \cite{Pineda:1997bj,Brambilla:1999xf}, 
when only ultrasoft
degrees of freedom remain dynamical. Such an effective 
field theory turns out to be, in fact, close to a Schr\"{o}dinger-like 
description of the bound state \cite{Bodwin:2006dn,Chung:2008sm}.
Moreover, matrix elements in pNRQCD can be expressed as the product
of WFOs and nonperturbative
glue-dependent factors, yielding a formal similarity with
potential models in many observable quantities (like decay widths
into leptons or light hadrons \cite{Brambilla:2002nu}, or
magnetic dipole transition \cite{Brambilla:2005zw}), 
which do not happen in NRQCD. 

The relation between
$\Lambda_{QCD}$ and the scales $m_Qv$ and $m_Qv^2$ dictates the
degrees of freedom of pNRQCD:
\begin{itemize}
\item The {\em weak coupling regime}, when $m_Qv^2 \gtrsim \Lambda_{QCD}
> m_Qv$
and the binding energy is mainly due to a Coulombic-like
potential. Dynamics can be described using perturbative theory.
\item The {\em strong coupling regime}, when $m_Qv \sim \Lambda_{QCD}$
and the binding energy is mainly due to
a confining (nonperturbative) potential.
\end{itemize} 
The assignment of each quarkonium state to any of these regimes is not 
such an easy task as the different scales are not directly measurable.
The fact that the spectrum of excitations of the bottomonium
family is not Coulombic suggests that the higher states are not in the weak
coupling regime. However, there are claims in the literature that the
$\Upsilon(2S)$ and $\Upsilon(3S)$ can also
be understood within the weak coupling regime 
\cite{Brambilla:2001qk,Brambilla:2001fw}. 
In Ref.\cite{GarciaiTormo:2005bs} $\Upsilon(1S,2S,3S)$
radiative decays were used to investigate their nature. 
Experimental results
from CLEO were confronted with the theoretical
expectations, in particular as a ratio of decay widths for
different energies of the photon.
 
In this Letter we follow a similar strategy but focusing on
electronic decays of the $\Upsilon(1S,2S,3S)$ 
resonances mainly basing our analysis on a phenomenological
approach. We have employed the
QQ-onia package presented in Ref.\cite{Domenech-QQ} to determine
the required WFOs, 
making use of a Cornell-type potential incorporating
a -$c'/r^2$ term as suggested by recent lattice studies. 
Table 1 shows the experimental values of
the electronic partial widths for all three resonances.

\begin{table*}[hbt]
\setlength{\tabcolsep}{0.6pc}
\caption{Measured electronic widths (in keV) for the
$\Upsilon(1S)$, $\Upsilon(2S)$ and $\Upsilon(3S)$ resonances 
(from \cite{pdg}).\newline}

\label{FACTORES}

\begin{center}
\begin{tabular}{ccccc}
\hline
$\Gamma_{ee}[\Upsilon(1S)]$  & $\Gamma_{ee}[\Upsilon(2S)]$ & 
$\Gamma_{ee}[\Upsilon(3S)]$ & \\
\hline
$1.340 \pm 0.018$ & $0.612 \pm 0.011$ & $0.443 \pm 0.008$ & \\
\hline
\end{tabular}
\end{center}
\end{table*}

\section{Velocity counting rules}

In this section we briefly review different velocity scaling rules appearing
in both perturbative and nonperturbative regimes of heavy quarkonia.
All our arguments are order-of-magnitude estimates which, moreover,
are subject to uncertainties especially in the nonperturbative regime
due to our ignorance of the scaling rules expected to be valid then. 

Notice that, no matter the regime, it always holds the following velocity
counting rules
\[ \langle nS| \frac{p^2}{m_b} |nS \rangle\ \sim\ m_bv^2\ ;\ 
\langle nS| V^{(0)} |nS \rangle\ \sim\ m_bv^2 \]
by the definition of $v$ and from the virial theorem, respectively.
The latter is an example where the naive guess of $m_bv$ is modified 
by the dynamics of the bound state to $m_bv^2$.

\subsection{Perturbative regime}

The soft scale in heavy quarkonium is basically set by its size $r$
which can be provided by the Bohr radius of the bound state.
If $r$  is small enough, i.e. 
$r \lesssim 1/\Lambda_{QCD}$,
the soft scale should be perturbative
and the potentials can be entirely determined in perturbation theory.

Then the following scaling rules {\em should} hold: 
\[ \langle nS| V^{(1)} |nS \rangle\ \sim \alpha_s^2/r^2 \sim m_b^2v^4 \]
since $\alpha_s \sim v$ and $r \sim 1/(m_bv)$, and
\[ \langle nS| V^{(2)} |nS \rangle\ \sim\ \alpha_s/r^3\ \sim m_b^3v^4 \]
and so on.

Let us stress that perturbation theory cannot incorporate quark confinement,
so it becomes crucial to determine the potential nonperturbatively
in this regime. 
This can be the case for high-lying quarkonium resonances, 
as we are checking in this work. Indeed, states below the $B\bar{B}$
threshold and not too deep (namely $\Upsilon(2S)$ and $\Upsilon(3S)$)
are expected to be in the strong-coupling regime whereas 
the deeper $\Upsilon(1S)$ state is expected to be in the weak coupling 
regime. States above (or very close to) open bottom threshold are
not expected to be in either regime \cite{GarciaiTormo:2007qs}.

\subsection{Nonperturbative regime}

Admittedly, the power counting of NRQCD is not well known 
in the nonperturbative regime and, in fact, has been addressed
by different authors in distinct ways
\cite{Brambilla:2002nu,Beneke:1997av,Fleming:2000ib}.
We might assume a very conservative counting:
$m_bv^d$ with $d$ standing for the operator 
dimension\footnote{This is somewhat similar to 
HQET \cite{Sanchis-Lozano:2001rr} where
any operator counts like $\Lambda_{QCD}^d$}. Thus
the following counting rules {\em could} hold:
\[ \langle nS | V^{(1)} |nS \rangle\ \sim\ m_b^2v^2\ ;\ 
\langle nS | V^{(2)} |nS \rangle\ \sim\ m_b^3v^3 \]

On the other hand, making use of the $1/m_b$ expansion
of the potential, one may write
\begin{equation}
V(r)=  V^{(0)}(r) + \frac{V^{(1)}(r)}{m_b}
+\frac{V^{(2)}(r)}{m_b^2}\ +\ \cdots
\end{equation}
where $V^{(k)}$, $k=0,1,2,\dots$ are the leading and subleading
terms respectively. The potential $V^{(2)}/m_b^2$ contains
the leading-order spin-dependent potentials and the
velocity-dependent potential. In this work we truncate the expansion
up to $V^{(1)}/m$, neglecting $1/m_b^2$ terms and higher.
In a nonperturbative regime this approximation should amount to
a ${\cal O}(v^2)$ accuracy at least when solving the Schr\"{o}dinger equation, 
thereby justifying the use of a non-relativistic approach.

Let us also point out that lattice calculations will be of help 
in determining the functional form of $V^{(1)}/m_b$.
We will come back to this important point in section 5.1.

Finally, notice 
that the static potential $V^{(0)}(r)$ is well parametrized by
a Coulomb plus linear term 
(i.e. a Cornell-type functional form \cite{Eichten:1978tg,Eichten:1979ms}),
\[ V^{(0)}(r)=-\frac{c}{r}+\sigma r + \mu \]
where $\sigma$ stands for the string tension governing the confining
potential and $\mu$ is a constant. This funnel shape 
will be used throughout this work as a reference, later on to be 
somewhat modified
when defining the actual leading-order potential for
heavy quarkonium.

\section{$\Upsilon$ leptonic decays}

As is well known, leptonic partial widths are a probe of the compactness
of the quarkonium system, and provide useful information complementary
to spectroscopy \cite{Rosner:2005eu}. In particular, the electronic width
for $^3S_1$ states ($\Gamma[\Upsilon(nS) \to e^+e^-]$) 
probes the WFO according to potential models.

Likewise, if the pNRQCD framework is applied, the leptonic
width can still be written in terms of the radial WFO
as mentioned in the Introduction. The following
expression obtained in Ref.\cite{Brambilla:2002nu} should hold up
to order $v^3\times(E_n/m_b,\Lambda_{QCD}^2/m_b^2)$:
\[
\Gamma[\Upsilon(nS) \to e^+e^-]\ =\ 
\frac{N_C}{\pi}\ \frac{|R_{n}(0)|^2}{m_b^2}\  
\times 
\]
\[
\biggl[\ \mathrm{Im} f_{ee}(^3S_1)\ \biggl(1-\frac{E_{n}^{\mathrm{(0)}}}{m_b}
\frac{2\epsilon_3}{9}+\frac{2\epsilon_3^{(2,EM)}}{3m_b^2}+
\frac{c_F^2B_1}{3m_b^2}\biggr) \]
\begin{equation} 
+\ \mathrm{Im} g_{ee}(^3S_1)\ \biggl(\frac{E_n^{(\mathrm{0})}}{m_b}-
\frac{\epsilon_1}{m_b^2}\biggr)\ \biggr]
\end{equation}
where $N_C=3$ is a colour factor  
and $E_n^{(0)}$is the (leading-order) bound-state energy; 
$\epsilon$ and $B$ stand for universal 
(i.e. flavour and state independent) nonperturbative 
parameters, 
which can be expressed in terms of gluonic field-strength
correlators \cite{Brambilla:2002nu}. 
They can be determined either by experimental data or by 
lattice simulation, but still their numerical values 
are quite uncertain.

The matching coefficients $\mathrm{Im} f_{ee}(^3S_1)$
and $\mathrm{Im} g_{ee}(^3S_1)$, corresponing to 
the $O_1(^3S_1)$ and $P_1(^3S_1)$ operators
of the NRQCD Lagrangian, are given at order $\alpha_s$
by the expressions \cite{Bodwin:1994jh,Luke:1997ys}
\begin{equation}
\mathrm{Im} f_{ee}(^3S_1) =  
\frac{1}{3}\pi Q^2\alpha^2
\biggl[1-\frac{16\alpha_s}{3\pi}+{\cal O}(\alpha_s^2)\biggr]  
\end{equation}
\begin{equation}
\mathrm{Im} g_{ee}(^3S_1) = -\frac{4}{9} 
\pi Q^2\alpha^2\biggl[1-\frac{8\alpha_s}{3\pi}+{\cal O}(\alpha_s^2)\biggr]
\end{equation}
where $|Q|=1/3$ for the bottom quark.

At lowest order, one recovers from (2) the well known formula
\cite{Royen} expressed in our notation as
\[
\Gamma[\Upsilon(nS) \to e^+e^-]\ =\ 
\frac{N_C\ \mathrm{Im} f_{ee}(^3S_1)}{\pi}
\ \frac{|R_{n}^{(\mathrm{0})}(0)|^2}{m_b^2}\  
\]
showing that the leptonic width of a quarkonium vector state 
is primarily sensitive to the square of its radial WFO,
though perturbative and nonperturbative corrections are large indeed. 
Yet the unknown parameters in Eq.(2) do not
allow its direct comparison with experiment.

\section{Testing the nature of heavy quarkonium}

In the leptonic width ratio of two $S$-wave states, however, 
several terms cancel out, leading to
\begin{equation}
\frac{\Gamma[\Upsilon(nS) \to e^+e^-]}{\Gamma[\Upsilon(rS) \to e^+e^-]}=
\frac{|R_{n}(0)|^2}{|R_{r}(0)|^2} 
\times [1+\delta_{nr}] 
\end{equation}
up to corrections of order ${\cal O}(v^q)$, 
where $q$ will be later determined in our analysis, 
providing an insight on the nature of heavy quarkonium and
a hint at the velocity scaling rules to be applied.

\begin{table*}[hbt]
\setlength{\tabcolsep}{0.4pc}
\caption{Values of the predicted 
and experimental mass (in GeV), 
WFO squared (or derivative) in GeV$^{3+2\ell}$, 
mean square radius (in fm) and typical quark velocity for the
$\Upsilon(1S,2S,3S,4S)$, $\chi_b(1P,2P)$ and $\Upsilon(1D)$ states 
when the improved Cornell-type (Corn-mod) potential of Eq.(10) is employed.
For comparison, we present in the first column 
the masses obtained using a Cornell (Corn) potential \cite{Eichten:1979ms}. 
\newline}

\label{FACTORES 3}

\begin{center}
\begin{tabular}{lcccccc}
\hline
Resonance & Mass (Corn) & Mass (Corn-mod) & Exp. &  
$|R_{n\ell}^{\ell'}(0)|^2$ & $\langle r^2 \rangle^{1/2}$ & 
$\langle v^2 \rangle$  \\
\hline
$\Upsilon(1S)$ & 9.4603 & 9.4603 & 9.4603  & $12.65$ & $0.23$ & $0.090$  \\
\hline
$\chi_b(1P)$ & 9.96 & 9.8929 & 9.9001  & $1.409$ & $0.40$ & $0.071$  \\
\hline
$\Upsilon(2S)$ & 10.05 & 10.0236 & 10.0233 &  $6.444$ & $0.51$ & $0.087$    \\
\hline
$\Upsilon(1D)$ & 10.20 & 10.1476 &  10.1622 & $0.562$ & 0.53 & $0.078$  \\
\hline
$\chi_b(2P)$ & 10.31 & 10.2729 & 10.2600  & $1.854$ & $0.63$ & $0.089$  \\
\hline
$\Upsilon(3S)$ & 10.40 & 10.3750 & 10.3552 & $5.404$ & $0.71$ & $0.103$  \\
\hline
$\Upsilon(4S)$ & 10.67 & 10.6477 &  10.5794 & $5.194$ & 0.88 & $0.120$  \\
\hline
\end{tabular}
\end{center}
\end{table*}

The correcting factor $\delta_{nr}$ is given by 
\[
\delta_{nr}=\biggl(\frac{\mathrm{Im} g_{ee}(^3S_1)}
{\mathrm{Im} f_{ee}(^3S_1)}-\frac{2\ \epsilon_3(2m_b)}{9}\biggr) \times
\biggl[\frac{E_n^{(0)}-E_r^{(0)}}{m_b}\biggr] 
\]
and using Eqs.(3-4) we get
\begin{equation} 
\delta_{nr} =
\biggl(r_{ee}+\frac{2\ \epsilon_3(2m_b)}{9}\biggr) \times
\biggl[\frac{E_r^{\mathrm{(0)}}-E_n^{\mathrm{(0)}}}{m_b}\biggr]
\end{equation}
where we have defined $r_{ee}$ up to 
${\cal O}(\alpha_s^2)$ corrections,
\[ r_{ee}=\frac{4}{3} \times \biggl[1+\frac{8\alpha_s(2m_b)}{3\pi}\biggr] \]
Now, taking into account that
$E_n=M_n-2m_b$ with
$M_n$ being the meson mass \cite{Gremm:1997dq}, 
we can safely use the following relation
\[ E_r^{(0)} - E_n^{(0)} \simeq E_r - E_n= M_r - M_n \]
Hence Eq.(6) can be rewritten as
\begin{equation}
\delta_{nr}=\biggl(r_{ee}+\frac{2\ \epsilon_3(2m_b)}{9}\biggr)\times
\biggl[\frac{M_r-M_n}{m_b}\biggr]
\end{equation}

On the other hand, $\epsilon_3(\mu)$ stands as the only
nonperturbative gluonic parameter in (6), 
all others cancelling out in the ratio (5) at the desired accuracy.
There is an experimental
determination of this long-distance parameter 
\footnote{Let us note a factor $N_C=3$ of difference between the definitions
of the gluonic parameter 
$\epsilon$ in \cite{Brambilla:2001xy} and $\epsilon_3$ in
\cite{Brambilla:2002nu}.}
in \cite{Brambilla:2001xy}:
\[ \epsilon_3(\mathrm{1\ GeV})=1.8^{+1.2}_{-0.7} \]
where the error bars are experimental only.
Additional theoretical uncertainties associated to subleading
operators in the power counting and perturbative expansion
are not taken into account in the above uncertainty.
 
Finally, we need to know $\epsilon_3$ at the bottomonium scale.
To this aim we can use the scale evolution 
\begin{equation}
\epsilon_3(\mu')=\epsilon_3(\mu)
+\frac{24C_F}{\beta_0}\ln{\frac{\alpha_s(\mu')}{\alpha_s(\mu)}}
\end{equation}
where $C_F=4/3$, $\beta_0=11C_A/3-4n_fT_F/3$ with 
$C_A=3$, $T_F=1/2$ and $n_f=5$ in our case.

Setting $\epsilon_3(1\ \mathrm{GeV})=1.8$, yields
$\epsilon_3(2m_b)\simeq 4.2$ which can be then used
as an input in Eq.(7).

\subsection{Lattice estimates of $V^{(1)}/m_b$ shape and size}

As commented previously, the $V^{(1)}/m_b$ potential has the
form $1/r^2$ relying on perturbation theory in the short-distance region. 
However, since the binding energy of the $b-\bar{b}$ system,
typically of order $m_bv^2$, can be similar or even smaller than
$\Lambda_{QCD}$ due to the non-relativistic nature
of the system, it is essential to determine the potential
nonperturbatively.  Monte Carlo simulations of lattice QCD 
provide a powerful tool for 
a nonperturbative determination of the 
potential \cite{Koma:2006si,Koma:2007jq}.

In a former analysis presented in 
Ref.\cite{Koma:2006si}, $V^{(1)}(r)/m_b$ was found to be comparable
with the Coulombic term of the static potential (i.e. 
$V^{(1)}/m_b \sim 1/r$) when 
applied to bottomonium states up to $r=0.6$ fm. 
Consequently, if $V^{(1)}(r)$ is
nonperturbative, the piece $V^{(1)}(r)/m$ in the potential 
should not be considered
as subleading with respect to $V^{(0)}(r)$. 

However, in a later study involving further long distance data up to
$r=0.9$ fm \cite{Koma:2007jq}, the same authors found
that the $1/r$ function was not supported by the fit, while 
the functional form $1/r^2$ with the linear term could fit the
data well. It is interesting to note that $V^{(1)}/m_b$ turns out to 
have the same functional form
as expected from perturbative theory.

The $V^{(1)}/m$ term cannot be neglected as compared 
to the static potential $V^{(0)}$ and 
has to be incorporated into the quarkonium potential
for the sake of coherence. Thus the leading-order potential  
$V^{{\mathrm{LO}}}$ should read
\begin{equation}
V^{{\mathrm{LO}}}=V^{(0)} + \frac{V^{(1)}}{m_b}
\end{equation}
yielding a potential of the form
\begin{equation}
V_{Corn-mod}(r)=-\frac{c}{r}-\frac{c'}{r^2}+\sigma r + \mu
\end{equation}
We will refer to (10) as a {\em Cornell-modified} potential, since the
functional form has been improved by the additional
$-c'/r^2$ piece; besides, the contribution
from the $V^{(1)}/m$ term also alters the value of 
$\sigma$ \cite{Koma:2007jq}.

\begin{table*}[hbt]
\caption{${\mid}R_{n}^{\mathrm{LO}}(0){\mid}^2$ (in GeV$^3$) valñues 
of $\Upsilon(ns)$ states ($n=1,2,3$) for several potentials:
Buchmuller-Tye ($m_b=4.88$ GeV) \cite{Eichten:1995ch}; 
Cornell \cite{Eichten:1995ch} ($m_b=5.18$ GeV);   
Cornell-modified (including a $-c'/r^2$ piece, $m_b=4.7$ GeV).
\newline}
\begin{center}
\begin{tabular}{|l|c|c|c|c|}    \hline 
Resonance &  ${\mid}R_{n}^{\mathrm{LO}}(0){\mid}^2_{B-T}$ & 
${\mid}R_n^{\mathrm{LO}}(0){\mid}^2_{Corn}$   
& ${\mid}R_n^{\mathrm{LO}}(0){\mid}^2_{Corn-mod}$ \\ 
\hline
$\Upsilon(1S)$ & $6.477$ & $14.07$ & $12.65$      \\
\hline
$\Upsilon(2S)$ & $3.234$ & $5.669$ & $6.444$  \\
\hline
$\Upsilon(3S)$ & $2.474$ & $4.271$ & $5.404$   \\
\hline
\end{tabular}
\end{center}
\end{table*}

In our approach the values of the parameters $m_b,c,c',\sigma$
and $\mu$ are obtained through a fitting procedure to the
bottomonium spectrum ($\Upsilon(1S)$ and $ \Upsilon(2S)$ states),
not from lattice estimates.
We obtain from the fit the following values 
for the parameters of the potential (10):
\[ \sigma=0.217\ {\mathrm GeV^2},\ c=0.400,\ c'=0.010\ \mathrm{GeV}^{-1}\]
and  $m_b=4.7$ GeV
\footnote{A significant coincidence is found between the
$m_b$ value used in the lattice calculation \cite{Koma:2007jq} 
and required in our fit.}.

The values of the predicted masses, WFOs 
and other properties of interest for different bottomonium states
using this potential 
are shown in Table 2. An excellent agreement with 
the experimental mass values of different resonances
in the spectrum can be observed. 

\subsection{Discussion}
The WFOs corresponding to the Cornell-modified potential 
were obtained using our code based on a Numerov technique
(See Ref.\cite{Domenech-QQ} for a thorough description of the
QQ-onia package.). 
A comparison of the WFOs of the 
$\Upsilon(1S,2S,3S)$ states
obtained using different potentials can be found in Table 3.

Let us remark that the inclusion of the $-c'/r^2$
term in the Cornell potential has a non-straightforward 
effect on the new resulting WFOs, for it implies a modification of both 
the Coulombic term and the $b$-quark mass obtained from the fit. Let
us also note that the $c'$ value is determined to a large extent 
by the $\Upsilon(2S)$ resonance in our fitting method 
(see \cite{Domenech-QQ} for more details).

On the other hand, as already pointed out in section 5.1,  
the perturbative calculation of $V^{(1)}/m_b$ yields the same
functional $r$-dependence as suggested by lattice studies. 
Therefore, one may look upon $V^{(1)}/m_b$ as an interpolating
term between the perturbative and non-perturbative regimes,
rendering the fit meaningful even using both $\Upsilon(1S,2S)$ states.
As a check of the fitting procedure (see Table 2), 
the predicted meson masses for the bottomonium family obtained from 
the Cornell-modified potential improve with respect to the
Cornell potential \cite{Eichten:1979ms}, when
compared with the experimental spectrum \cite{pdg}.
  
Moreover, the values of the WFOs shown in Table 3 
look self-consistent: the (absolute)
relative variations for the 
Cornell-modified versus the Cornell
potential are $\sim 10\%$, $14\%$ and $26\%$
for the $\Upsilon(1S,2S,3S)$ resonances, respectively. Indeed, one 
could naively expect a ${\cal O}(v^2)$ effect
in the perturbative regime as a consequence of
incorporating the new term into the potential, but increasingly larger 
variations for higher (thus dominantly non-perturbative) states. 

Finally, note that the $V^{(2)}/m_b^2$ (and higher) 
terms neglected in
the expansion (1) of the QCD potential should 
likely provide corrections of 
relative order ${\cal O}(v^2)$ to the WFOs  
obtained by solving the Schr\"{o}dinger equation 
with the leading-order potential (10). This counting is 
supported by the spin-dependent splitting 
experimentally found in the bottomonium spectrum.
In fact, if the static potential were exact, then the potential model
would reproduce QCD up to corrections of relative order $v^2$
\cite{Bodwin:2006dn}.

\section{Numerical results}

Now we proceed to check the validity of formula (5) by recasting it
onto the following double (experimental to theoretical) ratio:
\begin{equation}
\frac{\Gamma[\Upsilon(nS) \to ee]/\Gamma[\Upsilon(rS) \to ee]}
{(|R_{n}^{\mathrm{LO}}(0)|^2/|R_{r}^{\mathrm{LO}}(0)|^2) \times 
[1+\delta_{nr}]}=1+\Delta_{nr} 
\end{equation}
where we have introduced 
the dimensionless quantity $\Delta_{nr}\ (n,r=1,2,3,\ n \neq r$) parametrizing
the deviation from unity for different combinations of 
all three $\Upsilon(1S)$, $\Upsilon(2S)$, and $\Upsilon(3S)$ states.

The following experimental and theoretical inputs 
have been employed in our analysis:
\\

\noindent
{\bf 1)} The experimental input for the electronic widths can be
readily obtained from Table 1, 
allowing a determination of the ratios with relative error $\sim 1\%$

\[ \frac{\Gamma[\Upsilon(nS) \to e^+e^-]}
{\Gamma[\Upsilon(rS) \to e^+e^-]},\ \ \ (n,r=1,2,3\ \ n \neq r) \]

\noindent
{\bf 2)} The $|R_n^{\mathrm{LO}}(0)|^2$ values
for different potentials can be found in Table 3. We 
have assumed in (11) that
\[ \frac{|R_{n}(0)|^2}{|R_{r}(0)|^2}\ =\  
\frac{|R_{n}^{\mathrm{LO}}(0)|^2}{|R_{r}^{\mathrm{LO}}(0)|^2}
\ \times\ [1+{\cal O}(v^q)] \]
where $q$ is expected to be $2$ on account of the arguments given
in the previous section.\\

\noindent
{\bf 3)} $\delta_{nr}$ were computed according to 
Eq.(7) (valid up to order $(v^2,\alpha_s^2)$ corrections). 
Experimental meson masses and
$m_b$ values for each potential were used (see Table 3); 
we set $\epsilon_3(2m_b)=4.2$
derived from Eq.(8) using $\epsilon_3(\mathrm{1\ GeV})=1.8$
\cite{Brambilla:2001xy}.\\

\begin{table*}[hbt]
\setlength{\tabcolsep}{0.4pc}
\caption{$\Delta_{nr}$ (in $\%$) for different potentials
from Table 3 using $\epsilon_3(2m_b)=4.2$. Let us observe
that $\Delta_{23}$ always remains smaller than $\Delta_{12}$
and $\Delta_{13}$, as expected if both states $\Upsilon(2S,3S)$ 
were in the same (strong-coupling) regime.
\newline}

\label{FACTORES 2}

\begin{center}
\begin{tabular}{cccc}
\hline
Potential & B-T & Cornell & Cornell-modified \\
\hline
$\Delta_{13}$ & $20\%$ & $36\%$& $12\%$ \\
\hline
$\Delta_{12}$ & $15\%$ & $30\%$ & $14\%$ \\
\hline
$\Delta_{23}$ & $9\%$ & $10\%$ & $1\%$ \\
\hline
\end{tabular}
\end{center}
\end{table*}

Thus, if the two quarkonium states $n$ and $r$
were in the same (strong) regime, one should expect 
\begin{equation} 
\Delta_{nr}\ \lesssim\ {\cal O}(v^q)\ \sim\ 100 \cdot v^q(\%) 
\end{equation} 
Conversely, one should expect $\Delta_{nr} > 100 \cdot v^q(\%)$, 
if anyone of the states is in the strong-coupling regime and
the other in the weak-coupling regime. 

The values for $\Delta_{nr}$ obtained in our analysis 
for the Buchmuller-Tye, Cornell and Cornell-modified
potentials can be found in 
Table 4, representing our main result.

In particular, it turns out that when the Cornell potential
is employed we find that 
$\Delta_{13}$ is greater than $\Delta_{12}$, which in turn
is larger than $\Delta_{23}$, i.e. 
\[ \Delta_{13} \simeq 35\%\ \gtrsim \Delta_{12} \simeq 30\%
> \Delta_{23} \simeq 10\% \] 
in agreement with a counting rule providing $q=2$ for the latter case.

In sum, our results for the Cornell potential are thus consistent
with the expected level of accuracy (up to order $v^2$)
of Eq.(11), provided that both
$\Upsilon(2S)$ and $\Upsilon(3S)$ states belong to 
the strong-coupling regime, while the $\Upsilon(1S)$ state does not.

Furthermore, once the $-c'/r^2$ piece
is included in the Cornell-modified potential we get
\[ \Delta_{13} \simeq\ \Delta_{12} \simeq 10\%
> \Delta_{23} \simeq 1\% \] 
where $\Delta_{23}$ is now found to be remarkably small. 

Of course, there is an uncertainty coming from the $\epsilon_3$ value
in the computation of $\delta_{nr}$ according to Eq.(7).
Setting $\epsilon_3(2m_b)$ equal to zero, we find that 
$\Delta_{23}$ becomes appreciably worse. Thereby 
a non-null value of $\epsilon_3(2m_b)$ is clearly
favoured in our analysis. Demanding $\Delta_{23}=0$ in Eq.(11) we
get $\epsilon_3(2m_b)=3.2$ from Eq.(7) and, consequently,
$\epsilon_3(1\ \mathrm{GeV})=0.8$
using the running equation (8).

\section{Conclusions}

In this Letter we have presented a phenomenological study of
the electronic width ratios of 
$\Upsilon$ resonances (below open bottom production), 
finding evidence favouring both
$\Upsilon(2S)$ and $\Upsilon(3S)$ states in the strong coupling
regime, at the same time disfavouring the $\Upsilon(1S)$ in it,
in accordance with the conclusions from the analysis 
of Ref.\cite{GarciaiTormo:2005bs} based on radiative
decays of $\Upsilon$ resonances. 

Moreover, 
the agreement between the experimental and predicted 
$\Upsilon(2S)$/$\Upsilon(3S)$ ratios
is even better than naively expected from a conservative velocity counting
once the Cornell potential
becomes improved by a $-c'/r^2$ piece, motivated by recent 
lattice studies \cite{Koma:2006si,Koma:2007jq}. Therefore we can
conclude that our results (both from lepton widths and spectroscopy)
favour the inclusion of such a nonperturbative 
term into the Cornell potential.

Let us also point out that a value of the gluonic nonperturbative parameter:  
$\epsilon_3(1\ \mathrm{GeV}) \simeq\ 1.8^{+1.2}_{-0.7}$ 
(as found in \cite{Brambilla:2001xy}) is
compatible within errors 
with our analysis on leptonic decays. Actually, one might
turn the question round extracting $\epsilon_3(2m_b)$ from Eq.(11) 
using the experimental data from Table 1 and the Cornell-modified potential, 
yielding $\epsilon_3(2m_b)=3.2$ ($\epsilon_3(1\ \mathrm{GeV})=0.8$)
with an estimated uncertainty of $\sim 30\%$ 
assuming a $v^2 \sim 10\%$ accuracy
in Eq.(7). Additional theoretical uncertainties should increase
the allowed range though values of 
$\epsilon_3(1\ \mathrm{GeV})$ around unity are preferred
as a general result of our analysis. 

Finally, we want to stress the relevance of 
further accurate experimental
measurements of leptonic widths (among other properties) of heavy
quarkonia to carry out precise tests of effective theories of QCD
(likely useful to deal with nonperturbative effects
showing up at the LHC) and even direct searches for new physics 
\cite{SanchisLozano:2006gx,Fullana:2007uq}.
A future Super Flavour Factory would play an 
invaluable role in this regard.\\

{\noindent}
{\em Acknowledgments} We thank M. Koma, P. Gonzalez, J. Lee,
J. Soto, A. Vairo and J. Vijande for useful comments, and N. Brambilla
for a careful reading of the manuscript.

\end{document}